\documentclass[aps, prb, amssymb, amsmath, superscriptaddress, twocolumn] {revtex4-1}

\usepackage{chngcntr}

\usepackage{graphicx}
\usepackage{color}
\usepackage{amsmath}
\usepackage{enumitem}
\usepackage{amssymb}
\usepackage[hidelinks]{hyperref}
\usepackage{cancel}
\usepackage{ulem}%%%%%%%
\usepackage{multirow}
\newcommand{\be}{\begin{equation}}
	\newcommand{\ee}{\end{equation}}

\newcommand{\bea}{\begin{eqnarray}}
	\newcommand{\eea}{\end{eqnarray}}

\newcommand{\s}{\sigma}

\makeatother

\begin{document}
	\title{Generation of gauge magnetic fields in a kagome spin liqud candidate using the Dzyaloshinskii-Moriya interaction.}
	
\author{Byungmin Kang and Patrick A. Lee}
\affiliation{
Department of Physics, Massachusetts Institute of Technology, Cambridge, MA, USA
}
%\author{ Patrick A. Lee}
%\affiliation{
%Department of Physics and Institute for Quantum Information and Matter,
%California Institute of Technology, Pasadena, CA 91125, USA
%}

\begin{abstract}
The recent discovery of magnetization oscillations in a kagome spin liquid candidate motivates us to examine the origin of the gauge magnetic field term that can give rise to quantum oscillations of fermionic spinons. We find that in the presence of the Dzyaloshinskii-Moriya interaction and an average spin polarization, the spin permutation operator around the unit cell acquires an imaginary part, and a net gauge flux  is generated through the unit cell of the kagome lattice. This new mechanism of gauge field generation can account for the strength of the gauge magnetic field needed to explain the experiment.

\end{abstract} 

\maketitle

\section{Introduction}

 The quantum spin liquid is an exotic state of matter that has generated intense theoretical interests since the proposal by P. W. Anderson in 1973~\cite{ANDERSON1973}, and a great deal of effort has gone into  searching for  its realization in nature~\cite{khatua2023experimental}. While the initial proposal was for a system with antiferromagnetic interactions which fails to order due to quantum fluctuations in   frustrated lattices, it is now recognized that the spin liquid state is a prime example of the notion of emergence, where new degrees of freedom  which are absent in the microscopic Hamiltonian emerge at low energy and low temperature~\cite{Savary2016,zhou2017quantum}. For example, starting with a spin-1/2 Heisenberg model where excitations are $S=1$ spin flips, spinons which carry spin-1/2 and no charge emerge, together with an internal gauge field coupled with spinons. If the spinons are fermions, it may have a Fermi surface and the gauge field may be an $\mathrm{U}(1)$ gauge field. Much focus has been on two dimensional (2D) systems and the $\mathrm{U}(1)$ gauge field is a 2D version of the electromagnetic field in our world. The spinons coupled to the gauge magnetic field form Landau levels and may exhibit quantum oscillations. Indeed, a proposal was made by Motrunich~\cite{motrunich2006} that an insulator near the Mott  transition may be a spin liquid candidate and there is a linear coupling between the physical magnetic field perpendicular to the plane, $B_c$, and the gauge magnetic field $b$. Near the Mott transition, the ratio $\alpha=b/B_c$ was found to be be of order unity~\cite{motrunich2006}.

Recently Zheng et al.~\cite{zheng2023} reported magnetization oscillations in a kagome spin liquid candidate YCu$_3$(OH)$_6$Br$_2$[Br$_{1-y}$(OH)$_y$] (YCOB) which was interpreted as originating from  an emergent fermionic spinon coupled to the gauge magnetic field $b$. For a fixed magnetic field, The period of the oscillation is found to be proportional to $\cos(\theta)$ where $\theta$ is the angle between the applied $B$ field and the axis perpendicular to the kagome plane. This demonstrates the orbital origin of the effect. Furthermore, the analysis found that $\alpha$ is of the order of unity or even larger~\cite{zheng2023}. This large value of $\alpha$ is unexpected because YCOB is a robust insulator with a charge gap of several volts. In a Hubbard model described by hopping $t$ and repulsion $U$, the linear coupling between $b$ and $B_c$ found by Motrunich \cite{motrunich2006} is of order $t^3/U^2$. The quadratic restoring force is estimated to be of order the exchange energy $J=4t^2/U$ and we expect $\alpha$ to be of order $t/U$ and  small in the large-$U$ limit~\cite{zheng2023}. This motivates us to search for another mechanism for generating the gauge magnetic field.

\begin{figure}[h]
\includegraphics[width=0.45\textwidth]{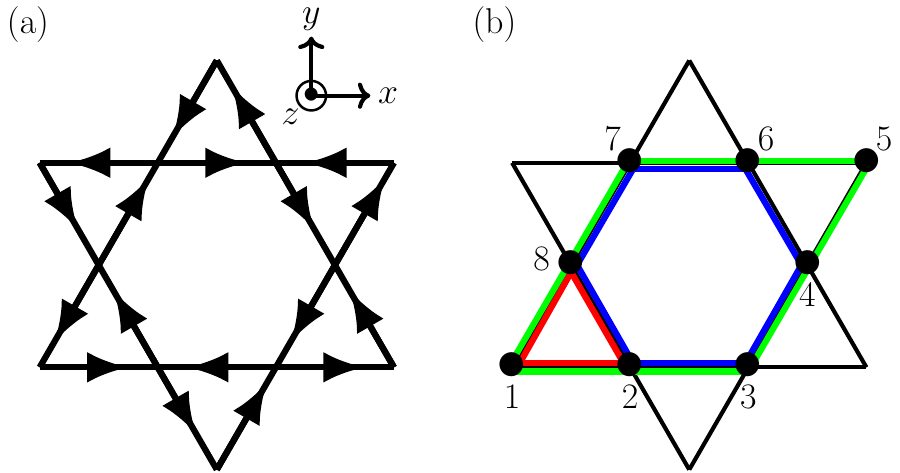}
\caption{(a) Convention of the ordering of vertices defining $\boldsymbol{D}_{i,j}$ in Eq.~\eqref{DM}. (b) Triangle loop, and the hexagon and kagome unit-cell are highlighted in red, blue, and greeen, respectively, used for the permutation operator. The sites in the perimeter of the kagome unit-cell is labelled in a counter-clockwsie order. } \label{fig_DM_cells}
\end{figure}

Generally speaking, the orbital signature of the oscillation suggests that spin-orbit coupling may be playing a role. In YCOB, as in the better known kagome system Herbertsmithite, there is a Dzyaloshiskii-Moriya (DM) interaction  in addition to the Heisenberg exchange term: 
\be \label{DM}
H
=  \sum_{\langle i,j \rangle} \big( J\, \boldsymbol{S}_i \cdot \boldsymbol{S}_j + \boldsymbol{D}_{i,j} \cdot \boldsymbol{S}_i \times \boldsymbol{S}_j \big),
\ee
where $J$ is the antiferromagnetic coupling constant and $\boldsymbol{D}_{i, j}$ is the DM vector. Let us focus on the $z$ component of the DM vector $\boldsymbol{D}_{i,j}$. The sign of the vector depends on the convention of ordering $i$ and $j$. This is indicated by arrows in Fig.~\ref{fig_DM_cells} (a), where we choose the convention that the arrows run counter-clockwise around each triangle in the kagome lattice, and we order $i$ and $j$ in Eq.~\eqref{DM} from the tail to the head of the arrow. With this convention, the $z$-component of the $\boldsymbol{D}_{i,j}$ vector, which we denote by $D_z$, are all equal in magnitude and  have the same sign~\cite{Elhajal2002}. It is convenient for us for work with the Pauli operators $\boldsymbol{\s}=2\boldsymbol{S}$ in the rest of the paper. For each triangle, the scalar chirality operator is defined by 
\be \label{chirality}
\hat{C}_{ijk}
=  \boldsymbol{\s}_i \cdot (\boldsymbol{\s}_j\times \boldsymbol{\s}_k),
\ee
where $i,j,k$ runs counter-clockwise around the triangle. In the presence of the DM interaction,  $D_0= \langle (\boldsymbol{\s}_j\times \boldsymbol{\s}_k)_z \rangle$ is nonzero. Lee and Nagaosa \cite{lee2013proposal}
made use of this fact to show that fluctuations in $\s_z$ couples linearly to fluctuations of the chirality operator, thus providing a way to measure chirality fluctuations through the triangles using neutron scattering.

Gao and Chen \cite{gao2020topological} took this one step further and argued that in the presence of a finite magentic field $B_c$  along the c-axis, $\langle \s_z \rangle$ is nonzero. Then the average chirality on the triangle $C_{123}=\langle \hat{C}_{123} \rangle \propto D_0 \langle \s_z \rangle$ is also nonzero. As shown by Wen, Wilzcek, and Zee~\cite{wen1989chiral}, the gauge magnetic flux through the triangle is proportional to the scalar chirality $C_{123}$. This is a mechanism to produce a gauge flux through the triangles. However, Gao and Chen stated that a negative gauge flux is generated through the hexagon in the kagome lattice which exactly cancels the flux through the triangles. As a result they concluded that there in no net flux through the unit cell. In their paper this conclusion was reached without providing any details. The aim of our work is to give a thorough derivation of the gauge flux through the unit cell. We reach the conclusion that there is finite gauge flux through the kagome unit cell that is proportional to $\langle S_z \rangle$. Hence the DM term can indeed generate the average gauge magnetic field $b$, that is needed to explain the observed magnetization oscillations.

\section{Spin permutation and chirality}
\label{sec:Spin permutation and chirality}

For a robust insulator, we would like to restrict ourselves to spin space and ignore any charge fluctuations. We begin by reviewing the various connections between operators in spin space to chirality and gauge flux as described by Wen, Wilzcek and Zee~\cite{wen1989chiral}. We introduce the spin permutation operator $\hat{P}_{1...n}$ which maps the state $\vert s_1, \ldots, s_n \rangle$ to $\vert s_n, s_1, \ldots, s_{n-1} \rangle$, i.e., shifts the spins forming a periodic array in an anti-clockwise order, where $s_i = \pm 1$ denotes the spin state on site $i$. The permutation operator can be decomposed into a product of pair exchanges $\hat{P}_{i,j}$ which in turn equals $\frac{1}{2} \big( 1 + \boldsymbol{\sigma}_i \cdot \boldsymbol{\sigma}_j \big)$. Thus, we arrive at an expression for the expectation value of the permutation operator
\begin{align} \label{permute}
&P_{1 \ldots n} = \langle \hat{P}_{1 \ldots n} \rangle \nonumber \\
&=  \frac{1}{2^{n-1}} \big\langle (1 + \boldsymbol{\sigma}_1 \cdot \boldsymbol{\sigma}_2) (1 + \boldsymbol{\sigma}_2 \cdot \boldsymbol{\sigma}_3) \ldots (1 + \boldsymbol{\sigma}_{n-1} \cdot \boldsymbol{\sigma}_n) \big\rangle .
\end{align}
Physically, the permutation operator describes the motion of a spin around a loop. If $P_{1 \ldots n}$ has an imaginary part, it means that the motion has picked up a Berry's phase which we can associate with the flux through the loop. By writing $P=|P|e^{i\Phi}$, we can compute $\Phi$ from the real and imaginary parts of $P$. In particular, our main aim is to show that the imaginary part of $P$ is nonzero for a loop that encloses a unit cell of the kagome lattice. We will identify $\Phi/2\pi$ as the gauge magnetic flux per unit cell   seen by the spinon  and calculate $\Phi$ to first order in the DM interaction. This calculation will form the core of this paper.

Before presenting the full results, we would like to mention another method of computing the gauge flux by Wen, Wilczek, and Zee~\cite{wen1989chiral}. They introduced a second operator $\hat{\chi}$ to describe the gauge flux, which is given by the product 
\be \label{chi}
\hat{\chi}_{1...n}
=  \hat{\chi}_{1,2} \hat{\chi}_{2,3}...\hat{\chi}_{n,1}
\ee
where
\be \label{chi22}
\hat{\chi}_{i,j}
=  f^\dagger_{i,\s}f_{j,\s}
\ee
and $f_{j,\s}$ is the annihilation operator of a fermion which satisfies the constraint that the occupation number is unity on each site. The operator $\hat{\chi}_{1 \ldots n}$ describes the motion of the fermion around a loop in the restricted subspace and its mean value $\chi_{1 \ldots n}=\langle \hat{\chi}_{1 \ldots n} \rangle$ can again be used to characterize the gauge flux seen by the spinon. We will return to discuss the second way of determining the gauge flux in Section \ref{sec:A second route to estimate the gauge flux.}.

\subsection{Permuation operator}
To see how the real and imaginary part of the expectation value of the loop operator $P_{1 \ldots n}$ Eq.~\eqref{permute} is related to Heisenberg and chirality terms, let us first consider the simplest example which is the permutation operator on three sites (triangle)~\cite{wen1989chiral}
\begin{align}
\hat{P}_{123} & = \frac{1}{4} (1 + \boldsymbol{\sigma_1} \cdot \boldsymbol{\sigma}_2) (1 + \boldsymbol{\sigma_2} \cdot \boldsymbol{\sigma}_3) \nonumber \\
& = \frac{1}{4} \big (1 + \boldsymbol{\sigma_1} \cdot \boldsymbol{\sigma}_2 + \boldsymbol{\sigma_2} \cdot \boldsymbol{\sigma}_3 + (\boldsymbol{\sigma_1} \cdot \boldsymbol{\sigma}_2) (\boldsymbol{\sigma_2} \cdot \boldsymbol{\sigma}_3) \big) \nonumber \\
&= \frac{1}{4} \big( 1 + \boldsymbol{\sigma_1} \cdot \boldsymbol{\sigma}_2 + \boldsymbol{\sigma_1} \cdot \boldsymbol{\sigma}_3 + \boldsymbol{\sigma_2} \cdot \boldsymbol{\sigma}_3 - i \hat{C}_{123} \big) .
\end{align}
The real part of $P_{123} = \langle \hat{P}_{123} \rangle$ is determined by the expectation value of the Heisenberg terms and the imaginary part is determined by the chiralities. Based on this observation, it seems natural to guess the formula for the expectation value of an arbitrary permutation operator as a sum of Heisenberg terms acting on all possible pairs of sites and a sum of chiralities acting on all possible triples of sites. However, such a naive guess fails immediately in the square case: 
\begin{align}
\hat{P}_{1234} = \frac{1}{8} \Big( &1 + \sum_{1 \le a < b \le 4} (\boldsymbol{\sigma_a} \cdot \boldsymbol{\sigma}_b) + ( \boldsymbol{\sigma_1} \cdot \boldsymbol{\sigma}_2) (\boldsymbol{\sigma_3} \cdot \boldsymbol{\sigma}_4) \nonumber \\
&+ ( \boldsymbol{\sigma_1} \cdot \boldsymbol{\sigma}_4) (\boldsymbol{\sigma_2} \cdot \boldsymbol{\sigma}_3) - ( \boldsymbol{\sigma_1} \cdot \boldsymbol{\sigma}_3) (\boldsymbol{\sigma_2} \cdot \boldsymbol{\sigma}_4) \nonumber \\
&- i \big( \hat{C}_{123} + \hat{C}_{124} + \hat{C}_{134} + \hat{C}_{234} \big) \Big) ,
\end{align}
where products of Heisenberg terms appear. So for larger $n$, the expression becomes more complicated and we find terms involving higher order in Heisenberg and chiralities in the imaginary part as well.

To find the correct expression for the permutation operator, we first recall the following identities of spin operators, where we defer the derivation to Appendix~\ref{app:eqs-derivation}:
\begin{align}
&(\boldsymbol{\sigma}_1 \cdot \boldsymbol{\sigma}_2) (\boldsymbol{\sigma}_2 \cdot \boldsymbol{\sigma}_3) = \boldsymbol{\sigma}_1 \cdot \boldsymbol{\sigma}_3 - i \hat{C}_{123} \label{eq:s-dot-s} \\
&i \hat{C}_{123} \, (\boldsymbol{\sigma}_3 \cdot \boldsymbol{\sigma}_4) = - (\boldsymbol{\sigma}_1 \cdot \boldsymbol{\sigma}_4) (\boldsymbol{\sigma}_2 \cdot \boldsymbol{\sigma}_3) \nonumber \\
&\qquad \qquad \qquad \quad + (\boldsymbol{\sigma}_1 \cdot \boldsymbol{\sigma}_3) (\boldsymbol{\sigma}_2 \cdot \boldsymbol{\sigma}_4) + i \hat{C}_{124}\label{eq:s-dot-c}
\end{align}
Then, let us note that the permutation operator $\hat{P}_{1 \ldots n}$ acting on $n$ sites is related to $\hat{P}_{1 \ldots n-1}$ via 
\be \label{P-n-from-n-1}
\hat{P}_{1 \ldots n} = \hat{P}_{1 \ldots n-1} \hat{P}_{n-1, n} = \hat{P}_{1 \ldots n-1} \frac{1 + \boldsymbol{\sigma}_{n-1} \cdot \boldsymbol{\sigma}_n}{2} .
\ee
Finally, starting from the expression for $\hat{P}_{1 \ldots n-1}$, the expression for $\hat{P}_{1 \ldots n}$ can be obtained from Eq.~\eqref{P-n-from-n-1} together with Eqs.~\eqref{eq:s-dot-s} and~\eqref{eq:s-dot-c}. The resulting expression of the permutation operator on 6 sites (hexagon) is given by
\begin{widetext}
\begin{align} \label{eq:hat-P-123456}
\hat{P}_{123456} = \frac{1}{32} \Big( &1 + \sum_{1 \le a < b \le 6} (\boldsymbol{\sigma_a} \cdot \boldsymbol{\sigma}_b) + \sideset{}{'}\sum_{\substack{(a,b),(c,d) \\ a < c}} \textrm{sign} (abcd) (\boldsymbol{\sigma_a} \cdot \boldsymbol{\sigma}_b) (\boldsymbol{\sigma_c} \cdot \boldsymbol{\sigma}_d) \nonumber \\
&+ \sideset{}{'}\sum_{\substack{(a,b),(c,d),(e,6) \\ b \le 3, \, 4 \le d < 5}} \textrm{sign} (abcde6) (\boldsymbol{\sigma_a} \cdot \boldsymbol{\sigma}_b) (\boldsymbol{\sigma_c} \cdot \boldsymbol{\sigma}_d) (\boldsymbol{\sigma_e} \cdot \boldsymbol{\sigma}_6) - \hat{C}_{123} \hat{C}_{456} \nonumber \\
&- i \sum_{1 \le a < b < c \le 6} \hat{C}_{a b c} -i \sideset{}{'}\sum_{\substack{(a,b),(c,d,e) \\ b \le 4 , \, b < d \\ a >= 4, \, a > e}} \textrm{sign}(abcde) (\boldsymbol{\sigma_a} \cdot \boldsymbol{\sigma}_b) \hat{C}_{cde} \Big) ,
\end{align}
where $\textrm{sign}(ab\ldots f)$ equals $(-1)$ to the power of the total number of exchanges in order to make the tuple $(a, b, \ldots, f)$ into an ascending order and $\sideset{}{'}\sum_{I_1, \ldots, I_k}$ for tuples $I_1=(a_1,\ldots,a_i), \ldots, I_k=(b_1,\ldots,b_j)$ denotes the summation over all possible assignments of $\{1,\ldots, n\}$ into tuples $I_1, \ldots, I_k$ satisfying the following rules: (1) the element of $\{1, \ldots, n\}$ appears at most once, (2) each tuple is in ascending order, and (3) satisfying any one set of the constraints specified in the summation. Each line denotes a different set of constraints to be satisfied and only one set of constraints (appearing in one of the line) needs to be satisfied. For example, $I_1=(1,2), I_2=(3, 4, 5)$ is a valid assignment in the last summation satisfying the first among two set of constraints in the sum while $I_1=(1,3), I_2=(3,2,4)$ is not since $I_2$ is not ordered and $3$ appears twice. Note also for example, $\sideset{}{'}\sum_{I_1=(a,b,c)} = \sum_{1 \le a < b < c \le n}$.
$P_{12345678}$ is given by
\begin{align} \label{eq:hat-P-12345678}
\hat{P}_{12345678} = \frac{1}{128} \Big( &1 + \sideset{}{'}\sum_{(a,b)} (\boldsymbol{\sigma_a} \cdot \boldsymbol{\sigma}_b) + \sideset{}{'}\sum_{\substack{(a,b),(c,d) \\ a < c}} \textrm{sign}(abcd) (\boldsymbol{\sigma_a} \cdot \boldsymbol{\sigma}_b) (\boldsymbol{\sigma_c} \cdot \boldsymbol{\sigma}_d) \nonumber \\
&+ \sideset{}{''}\sum_{\substack{(a,b),(c,d),(e,f) \\ b \le 5, \, b<d<f ,\, f \ge 6 ,\, a+c+e > 6}} \textrm{sign}(abcdef) (\boldsymbol{\sigma_a} \cdot \boldsymbol{\sigma}_b) (\boldsymbol{\sigma_c} \cdot \boldsymbol{\sigma}_d) (\boldsymbol{\sigma_e} \cdot \boldsymbol{\sigma}_f) \nonumber \\
& + \sideset{}{'}\sum_{\substack{(a,b),(c,d),(e,f),(g,8) \\ b \le 3, \, 4 \le d \le 5, \, 6 \le f \le 7}} \textrm{sign} (abcdefg8) (\boldsymbol{\sigma_a} \cdot \boldsymbol{\sigma}_b) (\boldsymbol{\sigma_c} \cdot \boldsymbol{\sigma}_d) (\boldsymbol{\sigma_e} \cdot \boldsymbol{\sigma}_f) (\boldsymbol{\sigma_g} \cdot \boldsymbol{\sigma}_8) - \sideset{}{'}\sum_{\substack{(a,b,c),(d,e,f) \\ c \le 5, \, d \ge 4, \, c < d}} \textrm{sign} (abcdef) \hat{C}_{abc} \hat{C}_{def} \nonumber \\ 
&- \sideset{}{'}\sum_{\substack{(a,b),(c,d,e),(f,g,h) \\ b \le 3, \, e = 5, \, h=8 \\ b \le 6 , \, e =3 , \, h=8 \\ a=7 ,\, e=3}} \textrm{sign}(abcdefgh) (\boldsymbol{\sigma_a} \cdot \boldsymbol{\sigma}_b) \hat{C}_{cde} \hat{C}_{fgh}  - i \sideset{}{'}\sum_{(a, b, c)} \hat{C}_{abc} -i \sideset{}{'}\sum_{\substack{(a,b),(c,d,e) \\ a \ge 4, \, a > e \\ b \le 6 ,\, b < d}} \textrm{sign} (abcde) (\boldsymbol{\sigma_a} \cdot \boldsymbol{\sigma}_b) \hat{C}_{cde} \nonumber \\
&-i \sideset{}{'}\sum_{\substack{(a,b),(c,d),(e,f,g) \\ b \le 4 \le d < f ,\, f \ge 6 ,\, a + c + e> 6 \\ b \le 4 \le f ,\, g \le 6 \le c \\ a, c \ge 4 \ge g,\, b < d}} \textrm{sign} (abcdefg) (\boldsymbol{\sigma_a} \cdot \boldsymbol{\sigma}_b) (\boldsymbol{\sigma_c} \cdot \boldsymbol{\sigma}_d) \hat{C}_{efg} \Big) ,
\end{align}
where the summation rule is the same as the one used in Eq.~\eqref{eq:hat-P-123456} except for $\sideset{}{''}\sum$ in the second line where we impose an additional constraint that when $b=5$, we consider only the tuples satisfying $b-a + d-c + f-e \le 11$ and $d-c+f-e \le 8$. Among those tuples, when $d-c+f-e \ge 6$, we only include tuples with $(f-8)(b-3)=0$ (so $f=8$ or $b=3$) and $\{a,c,e\} \ne \{2,3,4\}$ (as a set) and $d=7$ when $a=1$ in the summation. These additional rules in the summation reflects a highly non-trivial combinatorial nature of the loop operator expression.
\end{widetext}
As one can see, the permutation operator is expressed in terms of not only linear in Heisenberg and chirality terms but also higher-order in Heisenberg and chirality terms. Since the expectation value of the Heisenberg term is not small in the spin liquid system, it is important count higher-order terms properly. In the following, we present the mean-field approximation of the permutation operator expectation value.

\subsection{Mean-field approximation of permutation operator}
Here, we use mean-field approximation to evaluate the expectation value of the permutation operators. Even in the absence of magnetic order,  the nearest-neighbor Heisenberg term has a non-zero expectation value related to the ground state energy. We will make the approximation of keeping only the nearest-neighbor terms: 
\be \label{eq:mf-heisenberg}
\langle \boldsymbol{\sigma}_i \cdot \boldsymbol{\sigma}_j \rangle =\begin{cases}
S_0 \qquad \textrm{if $(i,j)$ are nearest-neighbor} \\
0 \qquad \quad \textrm{otherwise}
\end{cases}
\ee
Similarly, in the presence of the DM term, we keep only the nearest neighbor term
\be \label{eq:mf-sigma-times-sigma}
\langle (\boldsymbol{\sigma}_i \times \boldsymbol{\sigma}_j)_z \rangle = \textrm{sgn}(i,j) D_0
\ee
where $\textrm{sgn}(i,j)=1$ if $i,j$ is along the arrow in Fig~\ref{fig_DM_cells} (a), and $-1$ if it is opposed. Using these mean-field approximation, the expectation value of the chirality term for any mutually distinct sites $i,j,k $ is given by
%\begin{align} \label{chiral3}
%& \langle \hat{C}_{ijk} \rangle \approx \frac{1}{3} \Big( \langle (\boldsymbol{\sigma}_i)_z \rangle \langle (\boldsymbol{\sigma}_j \times \boldsymbol{\sigma}_k)_z + \langle (\boldsymbol{\sigma}_j)_z \rangle \langle (\boldsymbol{\sigma}_k \times \boldsymbol{\sigma}_i)_z \nonumber \\
%&\qquad \qquad \quad + \langle (\boldsymbol{\sigma}_k)_z \rangle \langle (\boldsymbol{\sigma}_i \times \boldsymbol{\sigma}_j)_z \rangle \Big) \nonumber \\
%& = \frac{h}{3} \Big( \langle (\boldsymbol{\sigma}_j \times \boldsymbol{\sigma}_k)_z + \langle (\boldsymbol{\sigma}_k \times \boldsymbol{\sigma}_i)_z + \langle (\boldsymbol{\sigma}_i \times \boldsymbol{\sigma}_j)_z \Big) ,
%\end{align}
\begin{align} \label{chiral3}
\langle \hat{C}_{ijk} \rangle &= \langle \epsilon_{abc} (\boldsymbol{\sigma}_i)_a (\boldsymbol{\sigma}_j)_b (\boldsymbol{\sigma}_k)_c \rangle \nonumber \\
 &= \langle (\boldsymbol{\sigma}_i)_z (\boldsymbol{\sigma}_j \times \boldsymbol{\sigma}_k)_z \rangle + \langle (\boldsymbol{\sigma}_j)_z (\boldsymbol{\sigma}_k \times \boldsymbol{\sigma}_i)_z \rangle \nonumber \\
&\quad + \langle (\boldsymbol{\sigma}_k)_z (\boldsymbol{\sigma}_i \times \boldsymbol{\sigma}_j)_z \rangle \nonumber \\
& \approx \langle (\boldsymbol{\sigma}_i)_z \rangle \langle (\boldsymbol{\sigma}_j \times \boldsymbol{\sigma}_k)_z \rangle + \langle (\boldsymbol{\sigma}_j)_z \rangle \langle (\boldsymbol{\sigma}_k \times \boldsymbol{\sigma}_i)_z \rangle \nonumber \\
&\quad + \langle (\boldsymbol{\sigma}_k)_z \langle \langle (\boldsymbol{\sigma}_i \times \boldsymbol{\sigma}_j)_z .
\end{align}
In the first two lines we used operator identities and the last line is a mean field factorization. We set $\langle (\boldsymbol{\sigma}_i)_z \rangle = h$ to be site independent. Note that $i,j,k$ can be any 3 sites, not just the equilateral triangle formed out of nearest neighbor. Only pairs of $j,k$ need to be nearest-neighbor in order for $\langle (\boldsymbol{\sigma}_j \times \boldsymbol{\sigma}_k)_z \rangle$ to be non-zero to contribute to Eq.~\eqref{chiral3}. To simplify the expression, let us introduce $C_0 = D_0 h$ in the following.

Using the mean-field approximation, the expectation values of the flux operators for triangle, hexagon, and kagome unit-cell described in Fig.~\ref{fig_DM_cells} (b) are given by
\be \label{triangle}
P_{123} = \frac{1}{4} \big( 1 + 3 S_0 - i 3 C_0 \big)
\ee
\begin{align} \label{eq:P-123456-mf}
P_{123456} = \frac{1}{32} \Big[ &\Big( 1 + 6 S_0 + 9 S_0^2 + 2 S_0^3 - 4 C_0^2 \Big) \nonumber \\
& + i \Big( 24 C_0 + 23 S_0 C_0 \Big) \Big]
\end{align}
\begin{align} \label{eq:P-12345678-mf}
&P_{12345678} = \frac{1}{128} \Big[ \Big( 1 + 10 S_0 + 29 S_0^2 + 22 S_0^3 + 2 S_0^4 \nonumber \\
&\qquad \qquad + 2 S_0 C_0^2 \Big) + i \big( 8 C_0 + 22 S_0 C_0 + 5 S_0^2 C_0 \big) \Big] .
\end{align}
%The method on how we derived the equations is explained in Appendix~\ref{app:eqs-derivation}.
The computation of the numerical coefficients is quite involved due to  the complicated rules in assigning the signs of the terms in Eqs.~\eqref{eq:hat-P-123456} and~\eqref{eq:hat-P-12345678}. This was done with a computer symbolic manipulation code. It is worth noting that the leading term (zeroth order in $\langle \boldsymbol{\sigma}_i \cdot \boldsymbol{\sigma}_j \rangle$) in the imaginary part of $P$ in Eqs.~\eqref{eq:hat-P-123456} and~\eqref{eq:hat-P-12345678} is relatively simple. It is the sum over $\hat{C}_{abc}$ where $a,b,c$ covers all sites in ascending order. This can be counted by hand by enumerating all allowed triangles and the mean field value is calculated using Eqs.~\eqref{eq:mf-sigma-times-sigma} and~\eqref{chiral3}. So it can readily be checked that $P_{12345678}$ indeed has an imaginary part.

\begin{figure}[h]
\includegraphics[width=0.4\textwidth]{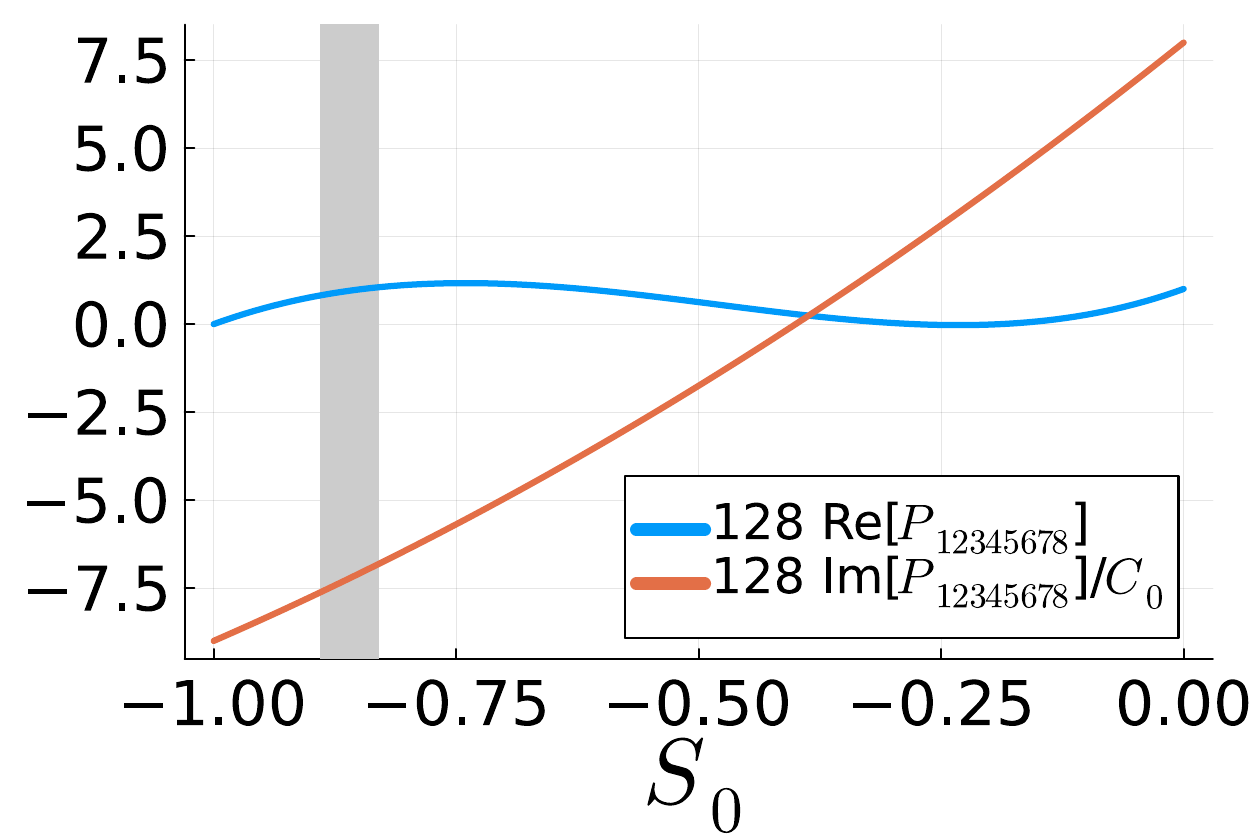}
\caption{Mean-field estimate of the real and imaginary part of the loop operator Eq.~\eqref{eq:P-12345678-mf} for the kagome unit-cell shown in Fig.~\ref{fig_DM_cells} (b) as functions of $S_0$. Note that we %rescale the $y$-axis by a linear factor
 have multiplied $P_{12345678}$ by the pre-factor $128$ in Eq.~\eqref{eq:P-12345678-mf} for  better readability. We ignore $S_0 C_0^2$ term in the real part since it is negligible compared to other terms, and plot the linear coefficient of $C_0$ for the imaginary part. We highlighted the region around $S_0 \approx -0.86$, which is the estimated value of $S_0$ for the Heisenberg model on the kagome lattice. In this region, the real part $\approx 1.0/128$ and the imaginary part $\approx -(7.2/128)C_0$.} \label{fig_p-loop}
\end{figure}

Now we provide an estimate for the parameter $S_0$. For the Heisenberg model on the kagome lattice, the ground state energy is close to $-0.43J$ per site~\cite {ran2007projected}. Since there are two bonds per site, the energy per bond is $J \langle \boldsymbol{S}_i \cdot \boldsymbol{S}_j \rangle \approx-0.215J$, hence $S_0 \approx -0.86$. Note this value is negative and close to unity in magnitude. Therefore it is necessary to keep higher orders in $S_0$ in our calculation. In fact for the case of the triangle given in Eq.~\eqref{triangle}, the real part of $P_{123}$ is negative. This reflects  the tendency to generating a $\pi$ flux due to frustration in the triangle. 

We finally compute the flux of the kagome unit-cell using Eq.~\eqref{eq:P-12345678-mf} and our estimates on $S_0$. As can be seen from Fig.~\ref{fig_p-loop}, the flux $\Phi_{1 \ldots 8}$ associated with the kagome unit-cell is given by $\Phi_{1 \ldots 8} \approx - \tan^{-1} (7.2 \, C_0)$, which is non-zero when $C_0 \ne 0$. Since $C_0$ is non-zero whenever the DM term exists, the kagome unit-cell experiences non-zero flux upon introducing the DM term. Estimate of the size of the flux for the specific case of YCOB will be given in section \ref{sec:Conclusion.}.

\section{A second route to estimate the gauge flux.}
\label{sec:A second route to estimate the gauge flux.}

 In this section we comment on the second route to estimating the gauge flux using the $\hat{\chi}$ operators. Let us consider $\chi_{1 \ldots n}=\langle \hat{\chi}_{1 \ldots n} \rangle$ where $\hat{\chi}_{1 \ldots n}$ was defined in Eq.~\eqref{chi} and examine its imaginary part. A common approximation is to factorize it into a product of $\chi_{i,j} = \langle \hat{\chi}_{i,j} \rangle$ around the loop, i.e., $\chi_{1 \ldots n} \approx \chi_{1,2} \chi_{2,3} \ldots \chi_{n,1}$. In the absence of the DM interaction, it is obvious that this product is real when the loop is around a unit cell, because  $\chi_{i,j}$ is equal to one that is translated by a lattice vector which appears in the product as complex conjugate. For example, in Fig.~\ref{fig_DM_cells} (b), $\chi_{2,3}=\chi_{6,5}=\chi^*_{5,6}$. Suppose we work to first order in the DM interaction and consider the correction to $\chi_{i,j}$. The correction terms are also the same under translation, so that $\chi_{i,j}$ remains equal under translation and the product $\chi_{1...n}$ remains real. To find the imaginary part we have to go beyond the factorization approximation. The next level of approximation is to keep a factor $\chi_2(i,j,k)$ defined as $\langle \hat{\chi}_{i,j} \hat{\chi}_{j,k} \rangle$ and factorize the rest into products of $\chi_{l,m}$. Now we can see that there is no longer any cancellation between pieces that were previously related by translation. As an example, let us compare $\chi_2(2,3,4)$ and $\chi_2(4,5,6)$ in Fig.~\ref{fig_DM_cells} (b). These two triangles have different geometry because sites 4 and 6 are nearest neighbors while sites 2 and 4 are not. Therefore the correction due to DM terms which live only on nearest neighbors will be different. Furthermore, the contributions of $\chi_2(4,5,6)$ and $\chi_2(8,1,2)$ which come from the two opposing triangles add because they are related by a 180 degrees rotation and are both counterclockwise in their ordering. Therefore in general  $\chi_{1...8}$ will have an imaginary part to leading order in the DM interaction. Within the slave particle mean field theory, it is possible to perform an explicit calculation using the mean field Hamiltonian using diagrammatic Greens function techniques. The steps are sketched in Appendix~\ref{app:chi2-greens-function}, but since our goal is to show that the imaginary part of $\chi_{1...8}$ is not cancelled, we will not pursue an explicit computation here.

 \section{Conclusion}
\label{sec:Conclusion.}

We now make some estimate of the gauge flux generated by the DM interaction in the specific case of YCOB. The gauge magnetic flux 
through the unit cell is given by $\Phi/2\pi$ which is given by the ratio of the imaginary part to the real part of $P_{12345678}$. Using the value $S_0=-0.86$  obtained earlier, we read from Fig.~\ref{fig_p-loop},
\be \label{Berry}
\Phi
\approx -7.2 C_0 \approx -7.2 \langle (\boldsymbol{\s}_2\times \boldsymbol{\s}_3)_z \rangle\langle \s_z \rangle
\ee
for small values of $C_0$. Starting from Eq.~\eqref{DM}  and treating the spins classically, we estimate the canting angle to be $\approx |D_z|/J $. Hence very roughly, we estimated   $\langle (\boldsymbol{\s}_2\times \boldsymbol{\s}_3)_z \rangle \approx -D_z/J$. Furthermore $ |D_z|/J \approx \Delta g/g$ where $\Delta g/g $ is the g factor anisotropy which is roughy 0.1 in YCOB. Zorko et al.~\cite{zorko2019negative} found by neutron scattering in a related compound which has anti-ferromagnetic order that the ordering is a 120 degree anti-chiral state, This  implies that $\langle (\boldsymbol{S}_2\times \boldsymbol{S}_3)_z \rangle$ tends to be negative and $D_z >0$. (Note that our convention for the sign of the DM term is opposite to that used in Refs.~\onlinecite{ Elhajal2002, zorko2019negative}.) Near the 1/9 plateau, $\langle \s_z \rangle \approx -1/9$ for field along the c-axis. Taken together we estimate
\be \label{flux}
\Phi/2\pi
\approx (7.2/2\pi)(D_z/J)\langle \s_z \rangle \approx -1.2 \times 10^{-2}.
\ee
The spinon couples to this negative gauge flux with a positive gauge charge. It is convenient to express the Berry phase in terms of an effective magnetic field $b$ so that 
$\Phi=2\pi \phi/\phi_0$ where $\phi=b A_0$, $A_0$ is the unit cell area and $\phi_0=h/e$ is the flux quantum. 
In YCOB, the unit cell area $A_0=38.53 \AA^2$ and we find that the effective magnetic field that gives this flux to be $b \approx 10^4 (\Phi/2\pi) $T which is about $-120$T. This is larger than the physical magnetic field $B \approx 30$T used in the experiment, so that $\alpha=|b|/B_c \approx 4$.  In Ref.~\onlinecite{zheng2023} $\alpha$ was found to be of order unity, but that estimate has large uncertainly because it depends quadratically on the assumed Dirac velocity which was not well determined. The important point is the flux generated by the DM interaction is large enough to give rise to the observed magnetization oscillations.
 
Up to now we estimated the DM contribution to the Berry phase  assuming that there is no other flux through the unit cell. In the case of YCOB we need to produce an extended unit cell with 9 bands in order to explain the 1/9 plateau. This can either come from breaking of translation symmetry, or by assuming $2\pi /3 $ flux per unit cell as was done in Ref.~\onlinecite{zheng2023}. This large flux produces the band structure with 9 bands, and the DM contribution should be considered as a small perturbation on this band structure. In particular, the flux we estimated in Eq.~\eqref{flux} gives an effective uniform gauge magnetic field $b$ which produces Landau levels in the bands near  the conduction and valence band edges and is the correct one to use to compare with the experiment. In principle, we should calculate the Berry phase using a tripled unit cell, where the model has a net flux of $2\pi $  and the hopping can be taken as periodic in the absence of DM. In practice we expect that Eq.~\eqref{flux} continues to be a reasonable estimate.

Finally, we note that the mechanism of generating a net gauge magnetic field from the DM term is quite general, and should be present as long as 
$\langle S_z \rangle$ is finite. For example, this will give rise to a thermal Hall effect even away from the 1/9 plateau if spinons are present. It should also be possible to use thermal Hall effect to probe the existence of spinons in other kagome systems which often have similar DM terms.
 
\section*{Acknowledgements} 

PL acknowledges support by DOE (USA) office of Basic Sciences Grant No. DE-FG02-03ER46076.

\bibliography{main}

%merlin.mbs apsrev4-1.bst 2010-07-25 4.21a (PWD, AO, DPC) hacked
%Control: key (0)
%Control: author (8) initials jnrlst
%Control: editor formatted (1) identically to author
%Control: production of article title (-1) disabled
%Control: page (0) single
%Control: year (1) truncated
%Control: production of eprint (0) enabled
\begin{thebibliography}{12}%
\makeatletter
\providecommand \@ifxundefined [1]{%
 \@ifx{#1\undefined}
}%
\providecommand \@ifnum [1]{%
 \ifnum #1\expandafter \@firstoftwo
 \else \expandafter \@secondoftwo
 \fi
}%
\providecommand \@ifx [1]{%
 \ifx #1\expandafter \@firstoftwo
 \else \expandafter \@secondoftwo
 \fi
}%
\providecommand \natexlab [1]{#1}%
\providecommand \enquote  [1]{``#1''}%
\providecommand \bibnamefont  [1]{#1}%
\providecommand \bibfnamefont [1]{#1}%
\providecommand \citenamefont [1]{#1}%
\providecommand \href@noop [0]{\@secondoftwo}%
\providecommand \href [0]{\begingroup \@sanitize@url \@href}%
\providecommand \@href[1]{\@@startlink{#1}\@@href}%
\providecommand \@@href[1]{\endgroup#1\@@endlink}%
\providecommand \@sanitize@url [0]{\catcode `\\12\catcode `\$12\catcode
  `\&12\catcode `\#12\catcode `\^12\catcode `\_12\catcode `\%12\relax}%
\providecommand \@@startlink[1]{}%
\providecommand \@@endlink[0]{}%
\providecommand \url  [0]{\begingroup\@sanitize@url \@url }%
\providecommand \@url [1]{\endgroup\@href {#1}{\urlprefix }}%
\providecommand \urlprefix  [0]{URL }%
\providecommand \Eprint [0]{\href }%
\providecommand \doibase [0]{http://dx.doi.org/}%
\providecommand \selectlanguage [0]{\@gobble}%
\providecommand \bibinfo  [0]{\@secondoftwo}%
\providecommand \bibfield  [0]{\@secondoftwo}%
\providecommand \translation [1]{[#1]}%
\providecommand \BibitemOpen [0]{}%
\providecommand \bibitemStop [0]{}%
\providecommand \bibitemNoStop [0]{.\EOS\space}%
\providecommand \EOS [0]{\spacefactor3000\relax}%
\providecommand \BibitemShut  [1]{\csname bibitem#1\endcsname}%
\let\auto@bib@innerbib\@empty
%</preamble>
\bibitem [{\citenamefont {Anderson}(1973)}]{ANDERSON1973}%
  \BibitemOpen
  \bibfield  {author} {\bibinfo {author} {\bibfnamefont {P.~W.}\ \bibnamefont
  {Anderson}},\ }\href@noop {} {\bibfield  {journal} {\bibinfo  {journal}
  {Materials Research Bulletin}\ }\textbf {\bibinfo {volume} {8}},\ \bibinfo
  {pages} {153} (\bibinfo {year} {1973})}\BibitemShut {NoStop}%
\bibitem [{\citenamefont {Khatua}\ \emph {et~al.}(2023)\citenamefont {Khatua},
  \citenamefont {Sana}, \citenamefont {Zorko}, \citenamefont {Gomil{\v{s}}ek},
  \citenamefont {Sethupathi}, \citenamefont {Rao}, \citenamefont {Baenitz},
  \citenamefont {Schmidt},\ and\ \citenamefont
  {Khuntia}}]{khatua2023experimental}%
  \BibitemOpen
  \bibfield  {author} {\bibinfo {author} {\bibfnamefont {J.}~\bibnamefont
  {Khatua}}, \bibinfo {author} {\bibfnamefont {B.}~\bibnamefont {Sana}},
  \bibinfo {author} {\bibfnamefont {A.}~\bibnamefont {Zorko}}, \bibinfo
  {author} {\bibfnamefont {M.}~\bibnamefont {Gomil{\v{s}}ek}}, \bibinfo
  {author} {\bibfnamefont {K.}~\bibnamefont {Sethupathi}}, \bibinfo {author}
  {\bibfnamefont {M.~R.}\ \bibnamefont {Rao}}, \bibinfo {author} {\bibfnamefont
  {M.}~\bibnamefont {Baenitz}}, \bibinfo {author} {\bibfnamefont
  {B.}~\bibnamefont {Schmidt}}, \ and\ \bibinfo {author} {\bibfnamefont
  {P.}~\bibnamefont {Khuntia}},\ }\href@noop {} {\bibfield  {journal} {\bibinfo
   {journal} {Physics Reports}\ }\textbf {\bibinfo {volume} {1041}},\ \bibinfo
  {pages} {1} (\bibinfo {year} {2023})}\BibitemShut {NoStop}%
\bibitem [{\citenamefont {Savary}\ and\ \citenamefont
  {Balents}(2016)}]{Savary2016}%
  \BibitemOpen
  \bibfield  {author} {\bibinfo {author} {\bibfnamefont {L.}~\bibnamefont
  {Savary}}\ and\ \bibinfo {author} {\bibfnamefont {L.}~\bibnamefont
  {Balents}},\ }\href {\doibase 10.1088/0034-4885/80/1/016502} {\bibfield
  {journal} {\bibinfo  {journal} {Reports on Progress in Physics}\ }\textbf
  {\bibinfo {volume} {80}},\ \bibinfo {pages} {016502} (\bibinfo {year}
  {2016})}\BibitemShut {NoStop}%
\bibitem [{\citenamefont {Zhou}\ \emph {et~al.}(2017)\citenamefont {Zhou},
  \citenamefont {Kanoda},\ and\ \citenamefont {Ng}}]{zhou2017quantum}%
  \BibitemOpen
  \bibfield  {author} {\bibinfo {author} {\bibfnamefont {Y.}~\bibnamefont
  {Zhou}}, \bibinfo {author} {\bibfnamefont {K.}~\bibnamefont {Kanoda}}, \ and\
  \bibinfo {author} {\bibfnamefont {T.-K.}\ \bibnamefont {Ng}},\ }\href@noop {}
  {\bibfield  {journal} {\bibinfo  {journal} {Reviews of Modern Physics}\
  }\textbf {\bibinfo {volume} {89}},\ \bibinfo {pages} {025003} (\bibinfo
  {year} {2017})}\BibitemShut {NoStop}%
\bibitem [{\citenamefont {Motrunich}(2006)}]{motrunich2006}%
  \BibitemOpen
  \bibfield  {author} {\bibinfo {author} {\bibfnamefont {O.~I.}\ \bibnamefont
  {Motrunich}},\ }\href@noop {} {\bibfield  {journal} {\bibinfo  {journal}
  {Physical Review B}\ }\textbf {\bibinfo {volume} {73}},\ \bibinfo {pages}
  {155115} (\bibinfo {year} {2006})}\BibitemShut {NoStop}%
\bibitem [{\citenamefont {Zheng}\ \emph {et~al.}(2023)\citenamefont {Zheng},
  \citenamefont {Zhu}, \citenamefont {Chen}, \citenamefont {Kang},
  \citenamefont {Zhang}, \citenamefont {Jenkins}, \citenamefont {Chan},
  \citenamefont {Zeng}, \citenamefont {Xu}, \citenamefont {Valenzuela} \emph
  {et~al.}}]{zheng2023}%
  \BibitemOpen
  \bibfield  {author} {\bibinfo {author} {\bibfnamefont {G.}~\bibnamefont
  {Zheng}}, \bibinfo {author} {\bibfnamefont {Y.}~\bibnamefont {Zhu}}, \bibinfo
  {author} {\bibfnamefont {K.-W.}\ \bibnamefont {Chen}}, \bibinfo {author}
  {\bibfnamefont {B.}~\bibnamefont {Kang}}, \bibinfo {author} {\bibfnamefont
  {D.}~\bibnamefont {Zhang}}, \bibinfo {author} {\bibfnamefont
  {K.}~\bibnamefont {Jenkins}}, \bibinfo {author} {\bibfnamefont
  {A.}~\bibnamefont {Chan}}, \bibinfo {author} {\bibfnamefont {Z.}~\bibnamefont
  {Zeng}}, \bibinfo {author} {\bibfnamefont {A.}~\bibnamefont {Xu}}, \bibinfo
  {author} {\bibfnamefont {O.~A.}\ \bibnamefont {Valenzuela}},  \emph
  {et~al.},\ }\href@noop {} {\bibfield  {journal} {\bibinfo  {journal} {arXiv
  preprint arXiv:2310.07989}\ } (\bibinfo {year} {2023})}\BibitemShut {NoStop}%
\bibitem [{\citenamefont {Elhajal}\ \emph {et~al.}(2002)\citenamefont
  {Elhajal}, \citenamefont {Canals},\ and\ \citenamefont
  {Lacroix}}]{Elhajal2002}%
  \BibitemOpen
  \bibfield  {author} {\bibinfo {author} {\bibfnamefont {M.}~\bibnamefont
  {Elhajal}}, \bibinfo {author} {\bibfnamefont {B.}~\bibnamefont {Canals}}, \
  and\ \bibinfo {author} {\bibfnamefont {C.}~\bibnamefont {Lacroix}},\
  }\href@noop {} {\bibfield  {journal} {\bibinfo  {journal} {Physical Review
  B}\ }\textbf {\bibinfo {volume} {66}},\ \bibinfo {pages} {014422} (\bibinfo
  {year} {2002})}\BibitemShut {NoStop}%
\bibitem [{\citenamefont {Lee}\ and\ \citenamefont
  {Nagaosa}(2013)}]{lee2013proposal}%
  \BibitemOpen
  \bibfield  {author} {\bibinfo {author} {\bibfnamefont {P.~A.}\ \bibnamefont
  {Lee}}\ and\ \bibinfo {author} {\bibfnamefont {N.}~\bibnamefont {Nagaosa}},\
  }\href@noop {} {\bibfield  {journal} {\bibinfo  {journal} {Physical Review
  B}\ }\textbf {\bibinfo {volume} {87}},\ \bibinfo {pages} {064423} (\bibinfo
  {year} {2013})}\BibitemShut {NoStop}%
\bibitem [{\citenamefont {Gao}\ and\ \citenamefont
  {Chen}(2020)}]{gao2020topological}%
  \BibitemOpen
  \bibfield  {author} {\bibinfo {author} {\bibfnamefont {Y.~H.}\ \bibnamefont
  {Gao}}\ and\ \bibinfo {author} {\bibfnamefont {G.}~\bibnamefont {Chen}},\
  }\href@noop {} {\bibfield  {journal} {\bibinfo  {journal} {SciPost Physics
  Core}\ }\textbf {\bibinfo {volume} {2}},\ \bibinfo {pages} {004} (\bibinfo
  {year} {2020})}\BibitemShut {NoStop}%
\bibitem [{\citenamefont {Wen}\ \emph {et~al.}(1989)\citenamefont {Wen},
  \citenamefont {Wilczek},\ and\ \citenamefont {Zee}}]{wen1989chiral}%
  \BibitemOpen
  \bibfield  {author} {\bibinfo {author} {\bibfnamefont {X.-G.}\ \bibnamefont
  {Wen}}, \bibinfo {author} {\bibfnamefont {F.}~\bibnamefont {Wilczek}}, \ and\
  \bibinfo {author} {\bibfnamefont {A.}~\bibnamefont {Zee}},\ }\href@noop {}
  {\bibfield  {journal} {\bibinfo  {journal} {Physical Review B}\ }\textbf
  {\bibinfo {volume} {39}},\ \bibinfo {pages} {11413} (\bibinfo {year}
  {1989})}\BibitemShut {NoStop}%
\bibitem [{\citenamefont {Ran}\ \emph {et~al.}(2007)\citenamefont {Ran},
  \citenamefont {Hermele}, \citenamefont {Lee},\ and\ \citenamefont
  {Wen}}]{ran2007projected}%
  \BibitemOpen
  \bibfield  {author} {\bibinfo {author} {\bibfnamefont {Y.}~\bibnamefont
  {Ran}}, \bibinfo {author} {\bibfnamefont {M.}~\bibnamefont {Hermele}},
  \bibinfo {author} {\bibfnamefont {P.~A.}\ \bibnamefont {Lee}}, \ and\
  \bibinfo {author} {\bibfnamefont {X.-G.}\ \bibnamefont {Wen}},\ }\href@noop
  {} {\bibfield  {journal} {\bibinfo  {journal} {Physical review letters}\
  }\textbf {\bibinfo {volume} {98}},\ \bibinfo {pages} {117205} (\bibinfo
  {year} {2007})}\BibitemShut {NoStop}%
\bibitem [{\citenamefont {Zorko}\ \emph {et~al.}(2019)\citenamefont {Zorko},
  \citenamefont {Pregelj}, \citenamefont {Gomil{\v{s}}ek}, \citenamefont
  {Klanj{\v{s}}ek}, \citenamefont {Zaharko}, \citenamefont {Sun},\ and\
  \citenamefont {Mi}}]{zorko2019negative}%
  \BibitemOpen
  \bibfield  {author} {\bibinfo {author} {\bibfnamefont {A.}~\bibnamefont
  {Zorko}}, \bibinfo {author} {\bibfnamefont {M.}~\bibnamefont {Pregelj}},
  \bibinfo {author} {\bibfnamefont {M.}~\bibnamefont {Gomil{\v{s}}ek}},
  \bibinfo {author} {\bibfnamefont {M.}~\bibnamefont {Klanj{\v{s}}ek}},
  \bibinfo {author} {\bibfnamefont {O.}~\bibnamefont {Zaharko}}, \bibinfo
  {author} {\bibfnamefont {W.}~\bibnamefont {Sun}}, \ and\ \bibinfo {author}
  {\bibfnamefont {J.-X.}\ \bibnamefont {Mi}},\ }\href@noop {} {\bibfield
  {journal} {\bibinfo  {journal} {Physical Review B}\ }\textbf {\bibinfo
  {volume} {100}},\ \bibinfo {pages} {144420} (\bibinfo {year}
  {2019})}\BibitemShut {NoStop}%
\end{thebibliography}%

\newpage

\appendix
\begin{widetext}
\section{Derivation of Eqs.~\eqref{eq:s-dot-s} and~\eqref{eq:s-dot-c}} \label{app:eqs-derivation}
In this section, we provide derivation of Eqs.~\eqref{eq:s-dot-s} and~\eqref{eq:s-dot-c} which played a central role in the main text. First of all, Eq.~\eqref{eq:s-dot-s} can be derived as 
\begin{align}
(\boldsymbol{\sigma}_1 \cdot \boldsymbol{\sigma}_2) (\boldsymbol{\sigma}_2 \cdot \boldsymbol{\sigma}_3) &= \sum_{a,b = x,y,z} (\boldsymbol{\sigma}_1)_a (\boldsymbol{\sigma}_2)_a (\boldsymbol{\sigma}_2)_b (\boldsymbol{\sigma}_3)_b = (\boldsymbol{\sigma}_1)_a (\boldsymbol{\sigma}_3)_b \Big( \delta_{ab} + i \epsilon_{abc} (\boldsymbol{\sigma}_2)_c \Big) \nonumber \\
&= \boldsymbol{\sigma}_1 \cdot \boldsymbol{\sigma}_3 - i \boldsymbol{\sigma}_1 \cdot (\boldsymbol{\sigma}_2 \times \boldsymbol{\sigma}_3) ,
\end{align}
where we adopt the Einstein summation convention in which repeated indices are summed over. Finally, Eq.~\eqref{eq:s-dot-c} can be derived as
\begin{align}
\big(i \boldsymbol{\sigma}_1 \cdot (\boldsymbol{\sigma}_2 \times \boldsymbol{\sigma}_3) \big) \, (\boldsymbol{\sigma}_3 \cdot \boldsymbol{\sigma}_4) &= i \epsilon_{abc} (\boldsymbol{\sigma}_1)_a (\boldsymbol{\sigma}_2)_b (\boldsymbol{\sigma}_3)_c (\boldsymbol{\sigma}_3)_d (\boldsymbol{\sigma}_4)_d = i \epsilon_{abc} (\boldsymbol{\sigma}_1)_a (\boldsymbol{\sigma}_2)_b \Big( \delta_{cd} + i \epsilon_{cde} (\boldsymbol{\sigma}_3)_e \Big) (\boldsymbol{\sigma}_4)_d \nonumber \\
&= - \epsilon_{abc} \epsilon_{cde} (\boldsymbol{\sigma}_1)_a (\boldsymbol{\sigma}_2)_b (\boldsymbol{\sigma}_3)_e (\boldsymbol{\sigma}_4)_d + i \boldsymbol{\sigma}_1 \cdot (\boldsymbol{\sigma}_2 \times \boldsymbol{\sigma}_4) \nonumber \\
&= - (\delta_{ad} \delta_{be} - \delta_{ae} \delta_{bd}) (\boldsymbol{\sigma}_1)_a (\boldsymbol{\sigma}_2)_b (\boldsymbol{\sigma}_3)_e (\boldsymbol{\sigma}_4)_d + i \boldsymbol{\sigma}_1 \cdot (\boldsymbol{\sigma}_2 \times \boldsymbol{\sigma}_4) \nonumber \\
&= -(\boldsymbol{\sigma}_1 \cdot \boldsymbol{\sigma}_4) (\boldsymbol{\sigma}_2 \cdot \boldsymbol{\sigma}_3) + (\boldsymbol{\sigma}_1 \cdot \boldsymbol{\sigma}_3) (\boldsymbol{\sigma}_2 \cdot \boldsymbol{\sigma}_4) + i \boldsymbol{\sigma}_1 \cdot (\boldsymbol{\sigma}_2 \times \boldsymbol{\sigma}_4) .
\end{align}
As explained in the main text, expressions for the permutation operator $\hat{P}_{1 \ldots n}$ is obtained via Eqs.~\eqref{eq:s-dot-s},~\eqref{eq:s-dot-c}, and~\eqref{P-n-from-n-1} starting from $n=2$ case. Since the number of terms grows exponentially in $n$, we used a computer symbolic manipulation code to enumerate all the terms, especially Eqs.~\eqref{eq:hat-P-123456} and~\eqref{eq:hat-P-12345678}. We then apply the mean-field approximation Eqs.~\eqref{eq:mf-heisenberg}--\eqref{chiral3} to get Eqs.~\eqref{triangle}--\eqref{eq:P-12345678-mf}, again with the aid of the computer program.

\section{Some details on $\chi_2 (i,j,k)$} \label{app:chi2-greens-function}
In this section we sketch a Green function diagrammatic method to calculate $\chi_2(i,j,k)= \langle \hat{\chi}_{i,j} \hat{\chi}_{j,k} \rangle$ to first order in the DM interaction. We consider only the $z$-component of the DM vector $D_{i,k}$ which we label as $D_z$. The DM contribution to the Hamiltonian can be written as
\be \label{HDM}
H_{\textrm{DM},z} = \sum_{\langle i,j \rangle}  D_z (\boldsymbol{S}_i \times \boldsymbol{S}_j)_ z
= \frac{iD_z}{2} \sum_{\langle i,j \rangle}\Big[ f^\dagger_{i,\uparrow}f_{i,\downarrow} f^\dagger_{j,\downarrow}f_{j,\uparrow} - \textrm{h.c.} \Big]
\ee
where we have used $2\boldsymbol{S}_i=f^\dagger_{i,\alpha} \boldsymbol{\sigma}_{\alpha,\beta} f_{i,\beta}$ and $\boldsymbol{\sigma}_{\alpha,\beta}$ are the Pauli matrices. Since $H_{\textrm{DM}, z}$ conserves the $z$-component of the total spin, the interaction vertex necessarily involves spin flips, which is represented in Fig.~\ref{fig_diagram} (a). Let us consider $\chi_2(4,5,6)$ to first order in $D_z$. It is given by the integral of the time ordered product
\begin{align} \label{chi2}
\chi_2(4,5,6) &= -i\int dt \langle T(\hat{\chi}_{4,5}(0)\hat{\chi}_{5,6}(0)H_{DM,z}(t) \rangle \nonumber \\
&=  -i^2 \frac{D_z}{2} \int dt \langle T[f^\dagger_{4,\alpha}(0)f_{5,\alpha}(0)f^\dagger_{5,\beta}(0)f_{6,\beta}(0)(f^\dagger_{i,\uparrow}(t)f_{i,\downarrow}(t) f^\dagger_{j,\downarrow}(t)f_{j,\uparrow}(t) - \textrm{h.c.}) \rangle] .
\end{align}
The vertices for the observable $\hat{\chi}_{4,5}(0)\hat{\chi}_{5,6}(0)$ are represented by the vertices in Fig.~\ref{fig_diagram} (b). The Feynman diagrams are generated by connecting the open lines in all possible ways. There are 4 kinds of diagrams shown in Fig.~\ref{fig_diagram} (c). (i) and (ii) are self energy corrections to the Green function due to the DM term. Note that a Hartree type correction is not allowed because a spin flip is necessary at the vertex. The self energy terms simply correct $\chi_{i,j}$ and will not contribute to an imaginary part to $P_{1...8}$, as explained earlier. The other diagrams are   (iii) and (iv). However, (iv) also violates the condition that spin flip is necessary at the interaction vertex and is forbidden. So the only diagram that contributes is (iii) and together with a similar diagram where all the spins are reversed and the interaction vertex changes sign. The diagrams are written in space and time; $(i,j)$ can be any pair, but it is a reasonable approximation to  limited them to the set $[4,5,6]$ and $t$ associated with the interaction vertex is integrated over. For a given mean field Hamiltonian, the unperturbed Green function $G_{\alpha,\beta}(i,j;t) = -i \langle T[f_{i,\alpha}(0)f^\dagger_{j,\beta}(t)] \rangle$  is diagonal in the spin index but depends on the spin in the presence of spin polarization. It can be evaluated and used to compute these diagrams. 

\begin{figure}
	\centering
	\includegraphics[width=0.7\textwidth]{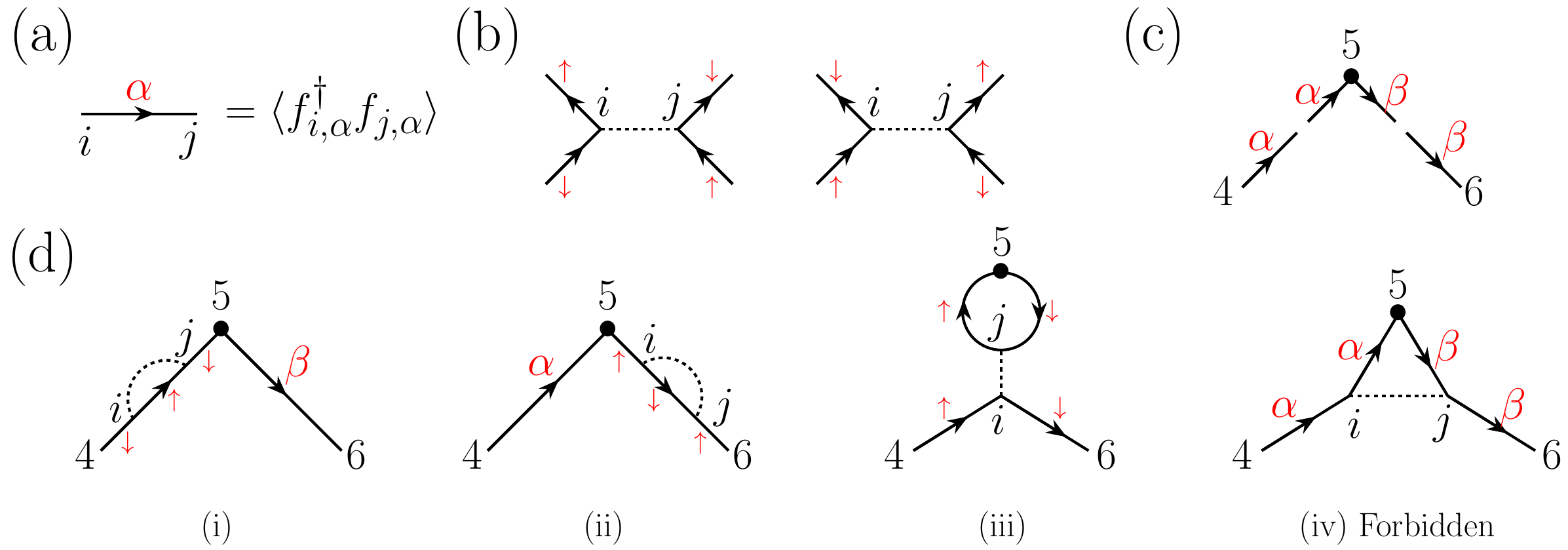}
	\caption{ 
Feynman diagrams for evaluating $\chi_2 (4,5,6)$. (a) $2$-point Greens function is represented by an arrow with site indices $i$ and $j$ and spin index $\alpha$. (b) Vertices represents the $z$-component of the DM interaction given by Eq.~\eqref{HDM}, which involves the spin flip at each site. (c) Vertices representing the operator product $\hat{\chi}_{4,5}(0)\hat{\chi}_{5,6}(0)$. (d) Diagrams that contribute to $\chi_2(4,5,6)=\langle \hat{\chi}_{4,5}(0) \hat{\chi}_{5,6}(0) \rangle$ to the first order in perturbation theory. Note that (iv) is forbidden since it cannot satisfy the spin-flip condition of the DM interaction.}
	\vspace{-2mm} \label{fig_diagram}
\end{figure}
\end{widetext}

\end{document}